\documentclass[preprint,showpacs,amsmath,nofootinbib]{revtex4}
\usepackage{mathrsfs}
\usepackage{graphicx}
\usepackage{amssymb}
\usepackage{amsmath}
\usepackage{color}

\definecolor{dyellow}{rgb}{1.,0.8,.0}
\definecolor{myblue}{rgb}{.1,.1,.7}
\definecolor{dcyan}{rgb}{.0,.6,.6}
\definecolor{cyan}{rgb}{0.4,1.0,1.0}
\definecolor{brown}{rgb}{0.6,0.2,0.}
\definecolor{darkblue}{rgb}{.0,.0,0.5}
\definecolor{darkred}{rgb}{0.75,0.0,0.0}
\definecolor{orange}{rgb}{1.,.6,.0}
\definecolor{dorange}{rgb}{0.8,.4,.0}
\definecolor{green}{rgb}{0.0,1.0,0.0}
\definecolor{darkgreen}{rgb}{0.0,0.6,0.0}
\definecolor{purple}{rgb}{.4,.0,.4}
\definecolor{lightgray}{rgb}{.7,.7,.7}
\begin{document}

\title{Critical Phenomena and Thermodynamic Geometry of RN-AdS Black Holes}

\author{Chao Niu and Yu Tian}
\affiliation{College of Physical Sciences, Graduate University of
Chinese Academy of Sciences, Beijing 100049, China}

\author{Xiao-Ning Wu}
\affiliation{Institute of Mathematics, Academy of Mathematics and
System Science, {CAS}, Beijing 100190, China} \affiliation{Hua
Loo-Keng Key Laboratory of Mathematics, {CAS}, Beijing 100190,
China}

\date{\today}

\begin{abstract}
The phase transition of Reissner-Nordstr\"om black holes in
$(n+1)$-dimensional anti-de Sitter spacetime is studied in details
using the thermodynamic analogy between a RN-AdS black hole and a
van der Waals liquid gas system. We first investigate critical
phenomena of the RN-AdS black hole. The critical exponents of
relevant thermodynamical quantities are evaluated. We find identical
exponents for a RN-AdS black hole and a Van der Waals liquid gas
system. This suggests a possible universality in the phase
transitions of these systems. We finally study the thermodynamic
behavior using the equilibrium thermodynamic state space geometry
and find that the scalar curvature diverges exactly at the van der
Waals-like critical point where the heat capacity at constant charge
of the black hole diverges.
\end{abstract}


\maketitle

%%%%%%%%%%%%%%%%%%%%%%%%%%%%%%%%%%%%%%%%%%%%%%%%%%%%%%%%%%%%%%%%
\section{Introduction}
Black hole is one of the most interesting objects in physics. The
study of black hole thermodynamics \cite{thermod1,thermod2} is therefore quite
important. Black holes are indeed thermodynamical objects with a
physical temperature and an entropy. It has been known over the past
few decades that thermodynamics of black holes provides an
important tool for understanding several issues involving quantum
theories of gravity. These have been intensely discussed for
the recent past. However, there is no microscopic or statistical
description behind their thermodynamical behavior, although
thermodynamic studies of black holes do indicate extremely rich
phase structures and critical phenomena in these systems. We can
consider black holes as states in a thermodynamical ensemble and to
study phase transition in black holes. A well-known example
is due to Hawking and Page \cite{hawk}. Motivated by these ideas, much work
has been done on phase structure of black holes, quite rich phase
structure and critical phenomena has been found \cite{critical}.

Recently, the study of phase transitions of black holes in
asymptotically anti de-Sitter spacetime
\cite{wu,cve1,cve2,Maeda1,sahay,ban} has focused much interest since
these transitions have been related with holographic
superconductivity \cite{gub,har,Maeda2} in the context of the
AdS/CFT correspondence (see relevant reviews in \cite{review}). In
this paper, we first review a thermodynamic analogy between a
$(n+1)$-dimensional RN-AdS black hole and a van der Waals liquid gas
system first discovered in \cite{myers,wu}. From this analogy we
calculate the critical exponents of relevant thermodynamical
quantities and discuss the scaling symmetry of the free energy. Then
we compare the critical exponents with the known case of a four
dimensional RN-AdS black hole \cite{wu} and a van der Waals liquid
gas system \cite{rei} and also check whether these exponents satisfy
the scaling law for the singular part of the free energy near
criticality. Among the results we get, we find that the critical
exponents of the four dimensional RN-AdS black hole in \cite{wu}
have some errors. But we find identical exponents between the
$(n+1)$-dimensional RN-AdS black hole and the van der Waals liquid
gas system, and possible universality in the phase transitions of
these systems. Furthermore we study the phase transition using a
geometrical perspective of equilibrium thermodynamics. This approach
has been developed over the last few decades
\cite{rupp1,rupp2,wei,sahay,ban}. We find the scalar curvature
diverges precisely at the van der Waals-like critical point where
the heat capacity at constant charge of the black hole diverges.

This paper is organized as follows. In section II we briefly discuss
the thermodynamics of the $(n+1)$-dimensional RN-AdS black
hole, mainly using it to establish our notations and obtain formulae
of thermodynamic functions for later use. In section III we study
the critical behavior of the $(n+1)$-dimensional RN-AdS black hole at
the van der Waals-like critical point. Further, in section IV, we
study the state space scalar curvature of the $(n+1)$-dimensional
RN-AdS black hole in detail. Finally, section V contain a discussion
of our results.

%%%%%%%%%%%%%%%%%%%%%%%%%%%%%%%%%%%%%%%%%%%%%%%%%%%%%%%%%%%%%%%%
\section{Phase Structure of a $(n+1)$-dimensional RN-AdS Black Hole}
The theme of the present Sec. is to give an overview of the
singular behavior of the heat capacity at constant charge of a
$(n+1)$-dimensional RN-AdS black hole which forms the background of
this work. For more details of the spherical case, see \cite{myers}.
Recently, motivated by the study of holographic superconductivity
\cite{gub,har}, plane symmetric and hyperbola symmetric cases in special
dimensions are also discussed \cite{sahay2}.

Now we consider general $(n+1)$-dimensional RN-AdS black holes. The
form of the spacetime metric is
\begin{equation}
ds^2=-f(r)dt^2+\frac{dr^2}{f(r)}+r^2d\Omega^2_{n-1},
\end{equation}
where\footnote{Throughout we shall adopt Planck units in which
$G=\hbar=c=k_B=1$, where all symbols have their usual meanings.}
\[f(r)=k-\frac{8\Gamma(\frac{n}{2})M}{(n-1)\pi^{\frac{n}{2}-1}r^{n-2}}+\frac{Q^2}{r^{2n-4}}+\frac{r^2}{l^2}\]
and $\Lambda=-\frac{n(n-1)}{2l^2}$ is the cosmological constant.
Here $k=1,0,-1$ corresponds to the sphere, plane and hyperbola symmetric cases respectively and
$d\Omega_{n-1}$ is the metric of the associated $(n-1)$-dimensional
base manifold.

The mass of the black hole is given by
\begin{equation}\label{mass}
M=\frac{(n-1)\pi^{\frac{n}{2}-1}}{8\Gamma(\frac{n}{2})}(kr^{n-2}_++\frac{Q^2}{r^{n-2}_+}+\frac{r_+^n}{l^2}),
\end{equation}
where $r_+$ is the value of $r$ at the horizon .

Using  the Bekenstein-Hawking formula, we have
\begin{equation}\label{entr}
S=\frac{A_n}{4}=\frac{\pi^{\frac{n}{2}}}{2\Gamma(\frac{n}{2})}r_+^{n-1}.
\end{equation}

It is now possible to determine the other thermodynamic entities
using the basic thermodynamic relations
\begin{equation}
\delta M=T\delta S+\Phi\delta Q.
\end{equation}
These are defined as
\begin{eqnarray}
T &= \Big(\frac{\partial M}{\partial S}\Big)_Q &= \frac{1}{4\pi} \frac{-\frac{2\Lambda}{n-1}r_+^{2n-2}+(n-2)kr_+^{2n-4}-(n-2)Q^2}{r_+^{2n-3}}\label{tem} \\
\Phi &= \Big(\frac{\partial M}{\partial Q}\Big)_S &= \frac{(n-1)\pi^{\frac{n}{2}-1}}{4\Gamma(\frac{n}{2})}\frac{Q}{r_+^{n-2}} \label{pot}\\
C_Q &= T\Big(\frac{\partial S}{\partial T}\Big)_Q &=
\frac{2(n-1)\pi^{\frac{n}{2}+1}}{\Gamma(\frac{n}{2})}\frac{r_+^{3n-4}T}{-\frac{2\Lambda}{n-1}r_
+^{2n-2}-(n-2)kr_+^{2n-4}+(n-2)(2n-3)Q^2}\label{hc}
\end{eqnarray}
where $ \Phi $ is the potential difference between the horizon and infinity,
$T$ is the Hawking temperature, $S$ is the entropy and $C_Q$ is the heat
capacity at constant charge of the black hole.

For a non-extreme black hole, it can be seen from (\ref{hc}) that $C_Q$ is
always positive and regular for the $k=0,-1$ cases, which tells us
that there is no phase transition happening. However, for the
spherically symmetry case $(k=1)$, $C_Q$ will become singular for a
certain set of black hole parameters $(M,Q)$ at which
\begin{equation}\label{sin}
-\frac{2\Lambda}{n-1}r_+^{2n-2}-(n-2)r_+^{2n-4}+(n-2)(2n-3)Q^2=0.
\end{equation}
Considering the equation (\ref{sin}), we then find that the critical points
are given in terms of the radius of the event horizon as $r_1, r_2
(r_1<r_2)$ when
$Q^2<(-\frac{(n-2)^2}{2\Lambda})^{n-2}\frac{1}{(n-1)(2n-3)}=:Q_c^2$. For the
special value
$Q^2=Q_c^2$, the two horizons degenerate, so we denote
$r_c:=r_1=r_2=\frac{n-2}{\sqrt{-2\Lambda}}$.

%%%%%%%%%%%%%%%%%%%%%%%%%%%%%%
\begin{figure}[h]
             \centering
             \includegraphics[scale=1]{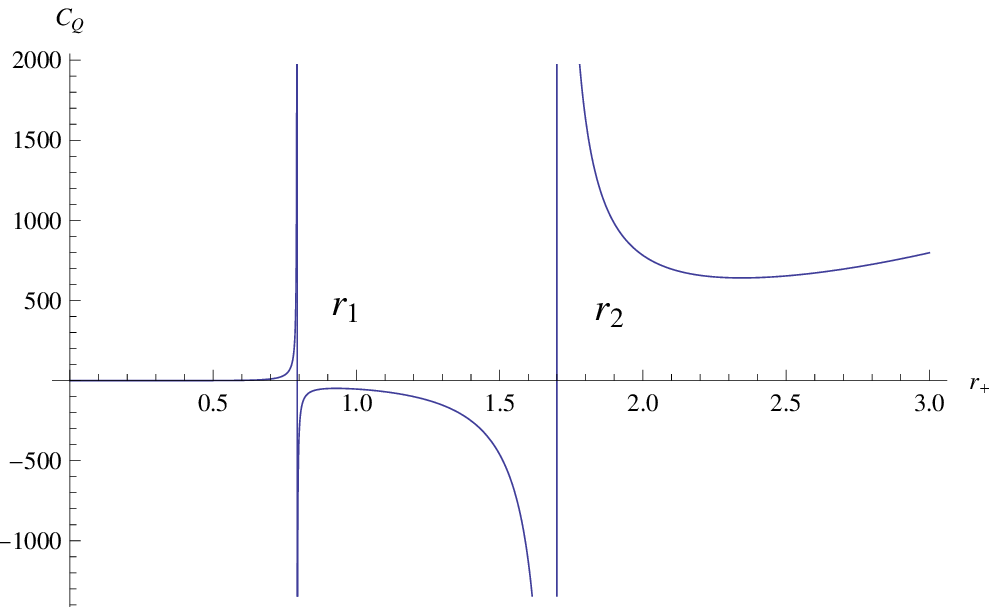}
             \caption{Heat capacity at constant charge with $r_+$ for $Q<Q_c$.}
             \label{fi1}
\end{figure}
%%%%%%%%%%%%%%%%%%%%%%%%%%%%%%

For fixed Q so that
$Q^2<Q_c^2$, $C_Q<0$ when $r_1<r_+<r_2$ and $C_Q>0$ when $r_+<r_1$
and $r_+>r_2$, so across the critical points at $r_1$ and $r_2$,
there is a change of thermodynamic stability of a black hole (see Fig.(\ref{fi1})).

%%%%%%%%%%%%%%%%%%%%%%%%%%%%%%
\begin{figure}[h]
             \centering
             \includegraphics[scale=1]{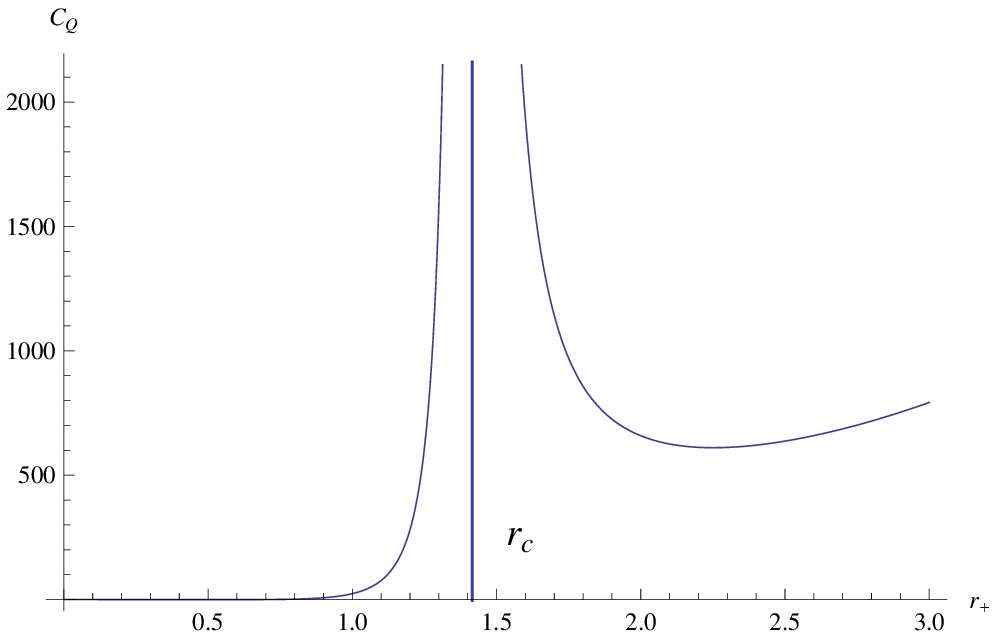}
             \caption{Heat capacity at constant charge with $r_+$ for $Q=Q_c$.}
             \label{fi2}
\end{figure}
%%%%%%%%%%%%%%%%%%%%%%%%%%%%%%

In the limit $Q^2$ approaches the critical value
$Q_c^2$, both $r_1$ and $r_2$ degenerate into $r_c$. In this case,
$C_Q$ remains positive and the unstable phase of a black hole
disappears (see Fig.(\ref{fi2})). When $Q^2$ is greater than $Q_c^2$, the
heat capacity $C_Q$ of the RN-AdS black hole is always regular.

%%%%%%%%%%%%%%%%%%%%%%%%%%%%%%%%%%%%%%%%%%%%%%%%%%%%%%%%%%%%%%%%
\section{Critical Behavior at the van der Waals-like critical point}

As described in II, when the charge of a RN-AdS black hole reaches
the critical value $Q_c$,the critical points at $r_1$ and $r_2$ degenerate
into a single critical point located at $r_c$.
The thermally unstable phase of a RN-AdS black hole disappears
(see Fig.(\ref{fi2})). The theme of this section is to study the critical
thermodynamic behavior of a RN-AdS black hole near $r_c$.
To this end, we shall first review a thermodynamic analogy
between a RN-AdS black hole and a van der Waals liquid gas
system. The analogy, though incomplete, will still serve as a
very useful guide in the study of the critical behavior of a
RN-AdS black hole in the vicinity of $r_c$.

\subsection{Thermodynamic analogy with a van der Waals liquid gas system}

Given the potential at the event horizon
$\Phi=\frac{(n-1)\pi^{\frac{n}{2}-1}}{4\Gamma(\frac{n}{2})}\frac{Q}{r_+^{n-2}}$,
the equation of state (\ref{tem}) can be rewritten as
\begin{equation}\label{state}
             T=\frac{1}{4\pi}\frac{(n-2)\Big(\frac{4\Gamma(\frac{n}{2})}{(n-1)\pi^{\frac{n}{2}-1}}\Phi\Big)^{\frac{2}{n-2}}
             -(n-2)\Big(\frac{4\Gamma(\frac{n}{2})}{(n-1)\pi^{\frac{n}{2}-1}}\Phi\Big)^{\frac{2n-2}{n-2}}-\frac{2\Lambda}{n-1}
             Q^{\frac{2}{n-2}}}{\Big(\frac{4\Gamma(\frac{n}{2})}{(n-1)\pi^{\frac{n}{2}-1}}Q\Phi\Big)^{\frac{1}{n-2}}}.
\end{equation}
In term of the thermodynamical variables $(Q,\Phi)$, we have
\begin{eqnarray}
             &&C_Q=\frac{(n-1)\pi^{\frac{n}{2}}}{2\Gamma(\frac{n}{2})}\times \nonumber \\
             &&\frac{(n-2)Q^{\frac{n-1}{n-2}}\Big(\frac{4\Gamma(\frac{n}{2})}{(n-1)\pi^{\frac{n}{2}-1}}\Phi\Big)^{\frac{2}{n-2}}
             -(n-2)Q^{\frac{n-1}{n-2}}\Big(\frac{4\Gamma(\frac{n}{2})}{(n-1)\pi^{\frac{n}{2}-1}}\Phi\Big)^{\frac{2n-2}{n-2}}
             -\frac{2\Lambda}{n-1}Q^{\frac{n+1}{n-2}}}
             {(n-2)(2n-3)\Big(\frac{4\Gamma(\frac{n}{2})}{(n-1)\pi^{\frac{n}{2}-1}}\Phi\Big)^{\frac{3n-3}{n-2}}
             -(n-2)\Big(\frac{4\Gamma(\frac{n}{2})}{(n-1)\pi^{\frac{n}{2}-1}}\Phi\Big)^{\frac{n+1}{n-2}}
             -\frac{2\Lambda}{n-1}Q^{\frac{2}{n-2}}
             \Big(\frac{4\Gamma(\frac{n}{2})}{(n-1)\pi^{\frac{n}{2}-1}}\Phi\Big)^{\frac{n-1}{n-2}}}\nonumber \\
\end{eqnarray}
and
\begin{equation}\label{par}
             \Big(\frac{\partial Q}{\partial\Phi}\Big)_T
             =\frac{Q}{\Phi}\frac{(n-2)(2n-3)\Big(\frac{4\Gamma(\frac{n}{2})}{(n-1)\pi^{\frac{n}{2}-1}}\Phi\Big)^{\frac{2n-2}{n-2}}
             -(n-2)\Big(\frac{4\Gamma(\frac{n}{2})}{(n-1)\pi^{\frac{n}{2}-1}}\Phi\Big)^{\frac{2}{n-2}}
             -\frac{2\Lambda}{n-1}Q^{\frac{2}{n-2}}}
             {(n-2)\Big(\frac{4\Gamma(\frac{n}{2})}{(n-1)\pi^{\frac{n}{2}-1}}\Phi\Big)^{\frac{2n-2}{n-2}}
             -(n-2)\Big(\frac{4\Gamma(\frac{n}{2})}{(n-1)\pi^{\frac{n}{2}-1}}\Phi\Big)^{\frac{2}{n-2}}
             -\frac{2\Lambda}{n-1}Q^{\frac{2}{n-2}}}.
\end{equation}

\begin{figure}[h]
             \centering
             \includegraphics[width=0.5\textwidth]{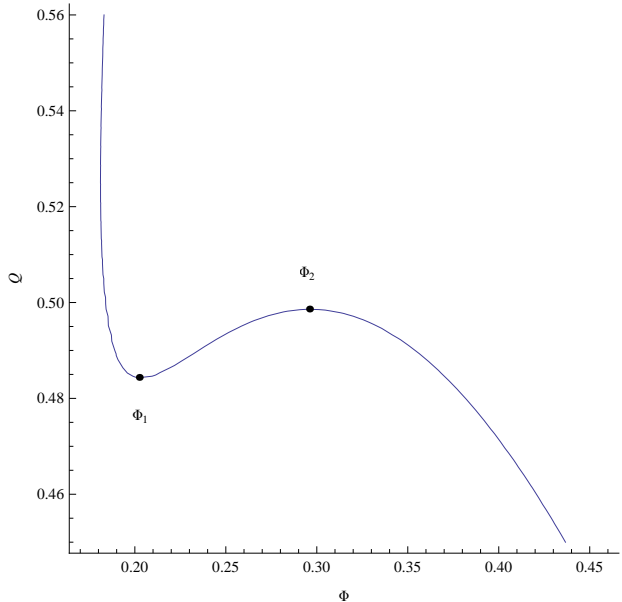}
             \caption{The isotherm of a RN-AdS black hole along which $T>T_c$. The local maxima and minima located respectively at $\Phi_1$ and $\Phi_2$ are critical points of $C_Q$. For $\Phi\in(\Phi_1,\Phi_2)$, the black hole is unstable with $(\frac{\partial Q}{\partial\Phi})_T>0$.}
             \label{fi3}
\end{figure}

It may be inferred from (\ref{par}) that, like a subcritical isotherm
of a van der Waals liquid gas system in the $(P,V)$ phase plane,
an isotherm of a RN-AdS black hole with $T>T_c$ also has a local
maxima and minima located respectively at $\Phi_1$ and $\Phi_2$.
Along the segment of the isotherm between $\Phi_1$ and $\Phi_2$,
a RN-AdS black hole is in a thermally unstable phase with
$(\frac{\partial Q}{\partial\Phi})_T>0$ (see Fig.(\ref{fi3})).

%%%%%%%%%%%%%%%%%%%%%%%%%%%%%%
%%%%%%%%%%%%%%%%%%%%%%%%%%%%%%

\begin{figure}[h]
             \centering
             \includegraphics[width=0.5\textwidth]{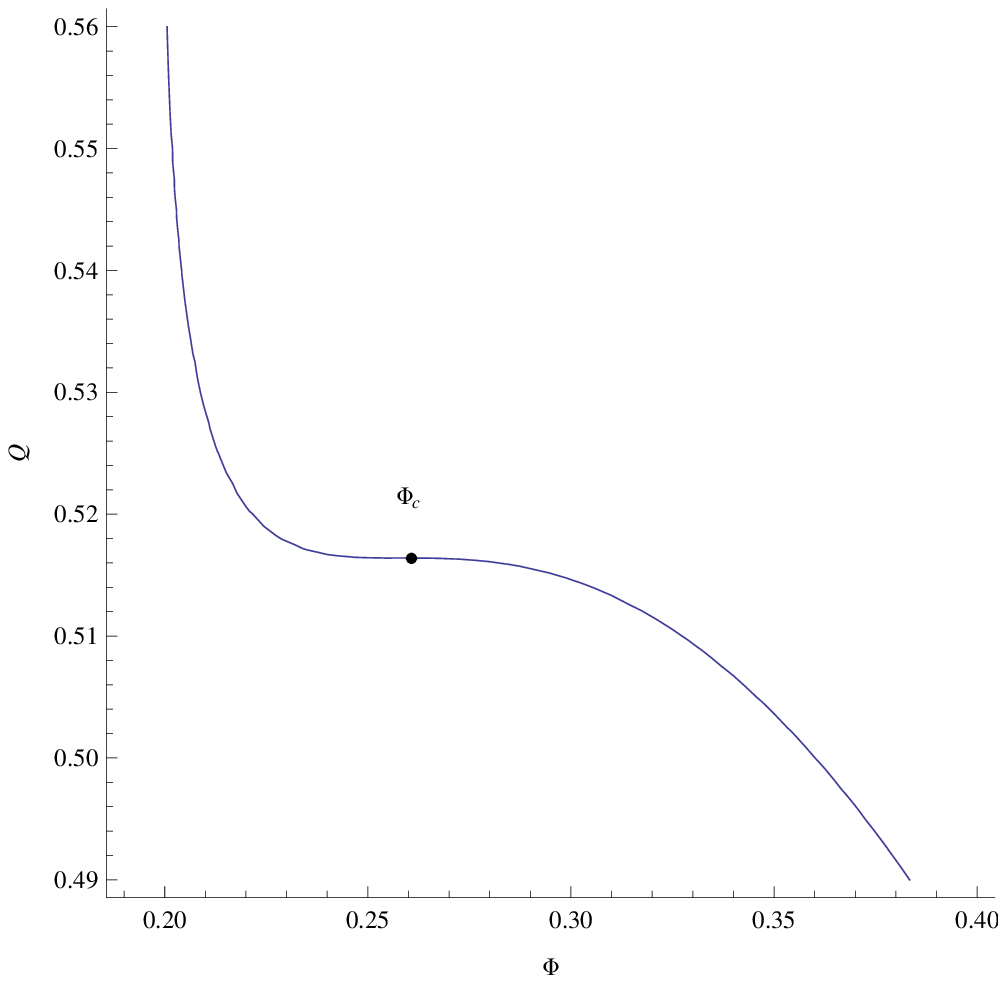}
             \caption{The critical isotherm along which $T=T_c$. The point of inflection located at $\Phi_c$ is a critical point of $C_Q$, $C_Q>0$ along the critical isotherm.}
             \label{fi4}
\end{figure}

In the limit when $T_c$ is reached, the shape of the isotherm
undergo noticeable change (see Fig.(\ref{fi4})) and the critical
points located at $\Phi_1$ and $\Phi_2$ on a subcritical isotherm
coalesce into a single critical point located at
$\Phi_c:=\frac{(n-1)\pi^{\frac{n}{2}-1}}{4\Gamma(\frac{n}{2})}\frac{Q_c}{r_c^{n-2}}
=\frac{(n-1)\pi^{\frac{n}{2}-1}}{4\Gamma(\frac{n}{2})}\frac{1}{\sqrt{(n-1)(2n-3)}}$
at the critical isotherm. The critical point at $\Phi_c$ coincides
with that located at $r_c$ on the critical isocharge curve with
$Q=Q_c$.

%%%%%%%%%%%%%%%%%%%%%%%%%%%%%%
%%%%%%%%%%%%%%%%%%%%%%%%%%%%%%

Like the case of the van der Waals liquid gas system, the
critical point at the critical isotherm (along which $T=T_c$)
of a RN-AdS black hole is also a point of inflection of the
critical isotherm and may be characterized by
\[\Big(\frac{\partial Q}{\partial\Phi}\Big)\Big|_c=0\]
\[\Big(\frac{\partial^2 Q}{\partial\Phi^2}\Big)\Big|_c=0\]
where the subscript $c$ denotes the corresponding quantity
evaluated at the critical point at $r_c$ from now on. In view
of the above similarities, if we formally identify the variables
$(Q,\Phi)$ of a RN-AdS black hole with $(P,V)$ of a van der Waals
liquid gas system, then we see that, at least at a qualitative level,
the phase structure of a RN-AdS black hole does bear certain
remarkable resemblances to that of a van der Waals liquid gas system.

\subsection{The introduction of an order parameter}

\begin{figure}[h]
             \centering
             \includegraphics[width=0.5\textwidth]{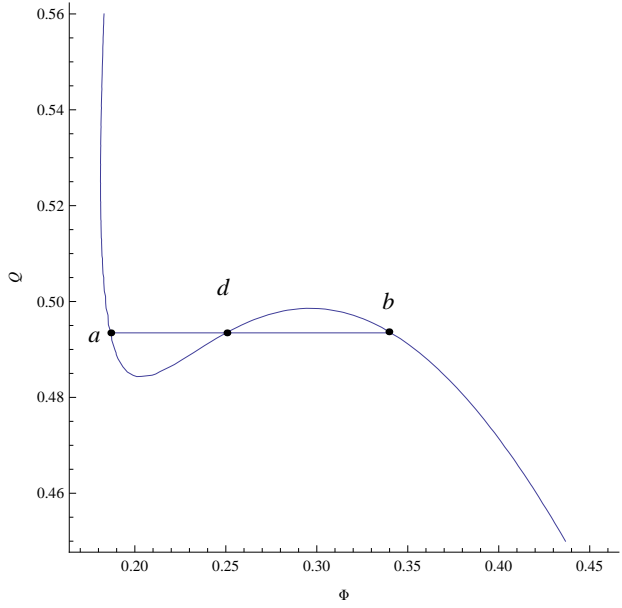}
             \caption{A horizontal line is drawn which connects
             points $a$ and $b$ of the subcritical isotherm. The
             area bounded by the line segment $ad$ and the
             isotherm is equal to that bounded by the line
             segment $db$ and the isotherm.}
             \label{fi5}
\end{figure}

In analogy to a van der Waals liquid gas system, an order
parameter in the RN-AdS context which measures the phase
change across the critical at $r_c$ may also be defined in
terms of the Maxwell equal-area law. To do so, in the $(Q,\Phi)$
phase plane, fix a subcritical isotherm and draw a horizontal
line which interests the subcritical isotherm at points $a,d,b$
(see Fig.(\ref{fi5})) such that the area bounded by the horizontal
line segment $ad$ and the isotherm is equal to that bounded
by the line segment $db$ and the isotherm.

As in the case of a van der Waals liquid gas system , define

\begin{equation}
             \eta=\Phi_b-\Phi_a
\end{equation}
as the order parameter to describe the phase change
of a RNAdS black hole near $r_c$.

\subsection{Critical exponents}

Near the critical point at the critical isotherm, the critical
behavior of a van der Waals liquid gas system may be
described in terms of
\[(1) \quad P-P_c\thicksim(V-V_c)^\delta\]
\[(2)\quad\frac{V_g-V_l}{V_c}\thicksim(-\epsilon)^\beta\]
\[ \begin{split}
             (3)\quad C_P&\thicksim(-\epsilon)^{-\alpha^{'}}\quad (T<T_c)\\
             &\thicksim\epsilon^{-\alpha}\quad (T>T_c)
\end{split} \]
\[ \begin{split}
             (4)\quad \kappa_T&\thicksim(-\epsilon)^{-\gamma^{'}}\quad (T<T_c)\\
             &\thicksim\epsilon^{-\gamma}\quad (T>T_c).
\end{split} \]
With the formal correspondence $(Q,\Phi)\leftrightarrow(P,V)$
as described in the preceding subsection, analogous quantities
may also be defined for a RN-AdS black hole. The concrete
values of the corresponding critical exponents in the case
of a RN-AdS black hole can also be worked out as follows.

\[1.\quad The~degree~of~the~critical~isotherm~\delta\]

Using the equation of state (\ref{state}), we have
\begin{eqnarray}\label{Qstate}
             &&Q^{\frac{1}{n-2}}=-\frac{n-1}{4\Lambda}
             \Bigg(4\pi\Big(\frac{4\Gamma(\frac{n}{2})}{(n-1)\pi^{\frac{n}{2}-1}}\Phi\Big)^{\frac{1}{n-2}}T-\nonumber \\ &&\sqrt{16\pi^2\Big(\frac{4\Gamma(\frac{n}{2})}{(n-1)\pi^{\frac{n}{2}-1}}\Phi\Big)^{\frac{2}{n-2}}T^2
             -\frac{8\Lambda(n-2)}{n-1}\bigg(\Big(\frac{4\Gamma(\frac{n}{2})}{(n-1)\pi^{\frac{n}{2}-1}}\Phi\Big)^{\frac{2n-2}{n-2}}
             -\Big(\frac{4\Gamma(\frac{n}{2})}{(n-1)\pi^{\frac{n}{2}-1}}\Phi\Big)^{\frac{2}{n-2}}\bigg)}\Bigg).\nonumber \\
\end{eqnarray}
In order to examine the neighborhood of the critical point, we
introduce expansion parameter $\epsilon =(T/T_c)-1$ and $\omega
=(\Phi/\Phi_c)-1$. In the neighborhood of the critical point, (\ref{Qstate})
can be written
\begin{equation}\label{exp}
             Q=a_{00}+a_{10}\epsilon+a_{01}\omega+a_{11}\epsilon\omega+a_{20}\epsilon^2+a_{02}\omega^2 +a_{21}\epsilon^2\omega+a_{12}\epsilon\omega^2+a_{30}\epsilon^3+a_{03}\omega^3+\cdots
\end{equation}
where $a_{\mu\nu}$ the coefficient of $\epsilon^\mu\omega^\nu$ in the expansion.

Let $\epsilon=0$ in (\ref{exp}). This gives
\begin{equation}
             Q=a_{00}+a_{01}\omega+a_{02}\omega^2+a_{03}\omega^3+\cdots
\end{equation}
where
\[a_{00}=Q_c\]
\[a_{01}=a_{02}=0\]
\[a_{03}\neq 0.\]
This means
\[\delta=3.\]

\[2.\quad The~degree~of~the~coexistence~curve~\beta\]

In the neighbourhood of the critical point, we have (\ref{exp}).
The values of $\omega$ on either side of the coexistence
curve can be found from the conditions that along the isotherm,
\begin{equation}\label{eq1}
\int_{\Phi_a}^{\Phi_b} \Phi dQ=0
\end{equation}
and
\begin{equation}\label{eq2}
Q(\Phi_a)=Q(\Phi_b).
\end{equation}
Let $\Phi_a=\Phi_c(1-\omega_a)$ and $\Phi_b=\Phi_c(1+\omega_b)$.
Substitute (\ref{exp}) into (\ref{eq1}) and (\ref{eq2}), we have
\begin{equation}\label{Eq1}
a_{11}\epsilon(\omega_b+\omega_a)+a_{21}\epsilon^2(\omega_b+\omega_a) +\frac{1}{2}(a_{11}+2a_{12})\epsilon(\omega_b^2-\omega_a^2)+a_{03}(\omega_b^3+\omega_a^3)=0
\end{equation}
and
\begin{equation}\label{Eq2}
a_{11}\epsilon(\omega_b+\omega_a)+a_{21}\epsilon^2(\omega_b+\omega_a) +a_{12}\epsilon(\omega_b^2-\omega_a^2)+a_{03}(\omega_b^3+\omega_a^3)=0.
\end{equation}
In order for (\ref{Eq1}) and (\ref{Eq2}) to be consistent, we must have $\omega_a=\omega_b$.
If we plug this into (\ref{Eq1}) or (\ref{Eq2}), we get $\omega_a=\omega_b=\omega$. This gives
\[\omega^2=-\frac{1}{a_{03}}(a_{11}\epsilon+a_{21}\epsilon^2).\]
Thus,
\[\omega_b \approx \omega_a=\sqrt{-\frac{a_{11}}{a_{03}}\epsilon}=(n-2)\sqrt{6\epsilon}.\]
This means
\[\beta=\frac{1}{2}.\]

\[3.\quad The~heat~capacity~exponent~\alpha\]

From (\ref{entr}), (\ref{tem}) and (\ref{pot}), we have
\begin{eqnarray}\label{hcp}
             C_\Phi=&&\frac{(n-1)\pi^{\frac{n}{2}}}{2\Gamma(\frac{n}{2})}\times \nonumber \\
             &&\frac{(n-2)Q^{\frac{n-1}{n-2}}\Big(\frac{4\Gamma(\frac{n}{2})}{(n-1)\pi^{\frac{n}{2}-1}}\Phi\Big)^{\frac{2}{n-2}}
             -(n-2)Q^{\frac{n-1}{n-2}}\Big(\frac{4\Gamma(\frac{n}{2})}{(n-1)\pi^{\frac{n}{2}-1}}\Phi\Big)^{\frac{2n-2}{n-2}}
             -\frac{2\Lambda}{n-1}Q^{\frac{n+1}{n-2}}}
             {(n-2)\Big(\frac{4\Gamma(\frac{n}{2})}{(n-1)\pi^{\frac{n}{2}-1}}\Phi\Big)^{\frac{3n-3}{n-2}}
             -(n-2)\Big(\frac{4\Gamma(\frac{n}{2})}{(n-1)\pi^{\frac{n}{2}-1}}\Phi\Big)^{\frac{n+1}{n-2}}
             -\frac{2\Lambda}{n-1}Q^{\frac{2}{n-2}}\Big(\frac{4\Gamma(\frac{n}{2})}{(n-1)\pi^{\frac{n}{2}-1}}\Phi\Big)
             ^{\frac{n-1}{n-2}}}.\nonumber \\
\end{eqnarray}
From (\ref{hcp}), we see that $C_\Phi$ display no singular behavior at the critical point. Therefore
\[\alpha=\alpha^{'}=0.\]

\[4.\quad The~isothermal~compressibility~exponent~\gamma\]

Let us compute $(\partial Q/\partial \omega)_\epsilon$. We obtain
\[\Big(\frac{\partial Q}{\partial\omega}\Big)_\epsilon=a_{11}\epsilon +a_{21}\epsilon^2+2a_{12}\epsilon\omega+3a_{03}\omega^2.\]
For $T<T_c$ one approaches the critical point along the critical isochore. Then set $\omega=0$, we obtain
\[\Big(\frac{\partial Q}{\partial\omega}\Big)_\epsilon=a_{11}\epsilon =2^{2-\frac{n}{2}}(n-2)^{n-2} \sqrt{(n-1)(2n-3)}(-\Lambda)^{1-\frac{n}{2}}\epsilon\]
for $\omega=0$. Therefore
\[\gamma^{'}=1.\]
For $T>T_c$ one approaches the critical point along the coexistence curve. Then set $\omega=\sqrt{-\frac{a_{11}}{a_{03}}\epsilon}$, we obtain
\[\Big(\frac{\partial Q}{\partial\omega}\Big)_\epsilon=-2a_{11}\epsilon =-2^{3-\frac{n}{2}}(n-2)^{n-2} \sqrt{(n-1)(2n-3)}(-\Lambda)^{1-\frac{n}{2}}\epsilon\]
for $\omega=\sqrt{-\frac{a_{11}}{a_{03}}\epsilon}$. Therefore
\[\gamma=1.\]

\subsection{Scaling symmetry for the Gibbs free energy near criticality}

In the case of a van der Waals liquid gas system, scaling symmetry
exists for the singular part of the Gibbs free energy near the
critical point located at the critical isotherm and the critical
exponents may all be expressed in terms of the two independent
homogeneity degrees of the Gibbs energy \cite{stan}. In this subsection, we
shall show that similar scaling symmetry also exists for a RN-AdS
black hole. The behavior of Gibbs free energy near the critical
point is similar to van der Waals system, from which scaling laws
for the critical exponents can be derived. Scaling symmetry in the
black hole critical phenomena was first discussed in \cite{lou} in the
context of of Kerr-Newman black holes.

Sufficiently close to $r_c$, the Gibbs free energy for a RN-AdS
black hole may be written as $G=G_r+G_s$. Here $G_r$ is the regular
part of the Gibbs free energy whose second order partial derivatives
are well behaved at the critical point at $r_c$, and $G_s$ is the
part of the Gibbs free energy responsible for the singular
thermodynamic behavior of a RN-AdS black hole near $r_c$. $G_s$ can
be further worked out to be
\begin{equation}\label{ge}
G_s=a\epsilon^2+b\omega^{4/3}
\end{equation}
for some constant $a,b$ dependent on $\Lambda$. From (\ref{ge}), we
find
\begin{equation}
G(\lambda^p\epsilon,\lambda^q\omega)=\lambda G(\epsilon,\omega)
\end{equation}
with $p=\frac{1}{2}, q=\frac{3}{4}$ and $\lambda$ a real
constant. As in the case of a van der Waals liquid-gas system, the
critical exponents derived in the previous section can be expressed
in terms of $p,q$ as
\begin{eqnarray}\label{sl1}
\alpha=&1-\frac{1}{p}\nonumber \\
\beta=&\frac{1-q}{p} \\
\gamma=&\frac{2q-1}{p}\nonumber \\
\delta=&\frac{q}{1-q}.\nonumber
\end{eqnarray}
From (\ref{sl1}), it may also be seen that the critical exponents in the
critical regime of $r_c$ are not independent. They are related by
following equations \cite{rei,stan}:
\begin{eqnarray}\label{sl2}
\alpha+2\beta+\gamma&=&2\nonumber \\
\alpha+\beta(\delta-1)&=&2 \\
\gamma(\delta-1)&=&(2-\alpha)(\delta-1)\nonumber \\
\gamma&=&\beta(\delta-1).\nonumber
\end{eqnarray}
Apart from obtaining the algebraic relations among the critical
exponents, (\ref{sl1}) or (\ref{sl2}) also enable us to give a
consistency check of the validity of the critical exponents
obtained in Sec. III C.

%%%%%%%%%%%%%%%%%%%%%%%%%%%%%%%%%%%%%%%%%%%%%%%%%%%%%%%%%%%%%%%%
\section{State Space Scalar Curvature for RN-AdS Black Hole}

In this section, we will study the critical phenomena of RN-AdS
black holes using thermodynamic geometry. The Hessian of the entropy
function (or other thermodynamical potentials) can be thought of as
a metric tensor on the state space \cite{rupp1,rupp2,wei,sahay,ban}. In the context of
thermodynamical fluctuation theory Ruppeiner has argued that the
Riemannian geometry of this metric gives insight into the underlying
statistical mechanical system. In this picture the occurrence of a
van der Waals critical point is connected with the divergence of the
state space scalar curvature. The metric as defined by Ruppeiner
\cite{rupp1} is given by,
\begin{equation}
g_{ij}=-\frac{\partial^2S(x)}{\partial x^i\partial x^j}
\end{equation}
where the coordinates $x^i$ are chosen to be the extensive variables of the
system. In fact, it is convenient to use the Weinhold metric which is defined
in the following way \cite{wei},
\begin{equation}
g^W_{ij}=-\frac{\partial^2U(x)}{\partial x^i\partial x^j}
\end{equation}
where we use $U$ to denote the internal energy. It
is well known that the line elements in Ruppeiner geometry and the
Weinhold geometry are conformally related by \cite{mru,sala}
\begin{equation}
dS^2_R=\frac{1}{T}dS^2_W
\end{equation}
where T is the temperature of the RN-AdS black hole. In this picture we
will consider $U=M-Q\Phi$, $x^1=S$ and $x^2=\Phi$.

From (\ref{mass}), (\ref{entr}), (\ref{tem}), (\ref{pot}) and (\ref{hc}), we have
\begin{eqnarray}
        M=&&\frac{(n-1)S^{\frac{n-2}{n-1}}}{4\pi}\Big(\frac{\pi^{\frac{n}{2}}}{2\Gamma(\frac{n}{2})}\Big)
        ^{\frac{1}{n-1}}\times\nonumber \\
        &&\bigg(1+\Big(\frac{4\Gamma(\frac{n}{2})}{(n-1)\pi^{\frac{n}{2}-1}}\Big)^2\Phi^2
        -\frac{2\Lambda}{n(n-1)}\Big(\frac{2\Gamma(\frac{n}{2})S}{\pi^{\frac{n}{2}}}\Big)^{\frac{2}{n-1}}\bigg)\label{mass2} \\
        T=&&\frac{1}{4\pi}\frac
        {-(n-2)\Big(\frac{4\Gamma(\frac{n}{2})}{(n-1)\pi^{\frac{n}{2}-1}}\Big)^2
        \Big(\frac{\pi^{\frac{n}{2}}}{2\Gamma(\frac{n}{2})}\Big)^{\frac{2}{n-1}}\Phi^2
        +(n-2)\Big(\frac{\pi^{\frac{n}{2}}}{2\Gamma(\frac{n}{2})}\Big)^{\frac{2}{n-1}}
        -\frac{2\Lambda}{n-1}S^{\frac{2}{n-1}}}
        {\Big(\frac{\pi^{\frac{n}{2}}S}{2\Gamma(\frac{n}{2})}\Big)^{\frac{1}{n-1}}}\label{tem2} \\
        C_Q=&&(n-1)S\times\nonumber \\
        &&\frac
        {-(n-2)\Big(\frac{4\Gamma(\frac{n}{2})}{(n-1)\pi^{\frac{n}{2}-1}}\Big)^2
        \Big(\frac{\pi^{\frac{n}{2}}}{2\Gamma(\frac{n}{2})}\Big)^{\frac{2}{n-1}}\Phi^2
        +(n-2)\Big(\frac{\pi^{\frac{n}{2}}}{2\Gamma(\frac{n}{2})}\Big)^{\frac{2}{n-1}}
        -\frac{2\Lambda}{n-1}S^{\frac{2}{n-1}}}
        {(n-2)(2n-3)\Big(\frac{4\Gamma(\frac{n}{2})}{(n-1)\pi^{\frac{n}{2}-1}}\Big)^2
        \Big(\frac{\pi^{\frac{n}{2}}}{2\Gamma(\frac{n}{2})}\Big)^{\frac{2}{n-1}}\Phi^2
        -(n-2)\Big(\frac{\pi^{\frac{n}{2}}}{2\Gamma(\frac{n}{2})}\Big)^{\frac{2}{n-1}}
        -\frac{2\Lambda}{n-1}S^{\frac{2}{n-1}}}.\nonumber\label{hc2} \\
\end{eqnarray}
Now using (\ref{mass2}) and (\ref{tem2})
we can easily calculate the Ruppeiner metric
\begin{eqnarray}
        g_{SS}=&&-\frac{1}{(n-1)S}\times\nonumber \\
        &&\frac
        {-(n-2)\Big(\frac{4\Gamma(\frac{n}{2})}{(n-1)\pi^{\frac{n}{2}-1}}\Big)^2
        \Big(\frac{\pi^{\frac{n}{2}}}{2\Gamma(\frac{n}{2})}\Big)^{\frac{2}{n-1}}\Phi^2
        +(n-2)\Big(\frac{\pi^{\frac{n}{2}}}{2\Gamma(\frac{n}{2})}\Big)^{\frac{2}{n-1}}
        +\frac{2\Lambda}{n-1}S^{\frac{2}{n-1}}}
        {-(n-2)\Big(\frac{4\Gamma(\frac{n}{2})}{(n-1)\pi^{\frac{n}{2}-1}}\Big)^2
        \Big(\frac{\pi^{\frac{n}{2}}}{2\Gamma(\frac{n}{2})}\Big)^{\frac{2}{n-1}}\Phi^2
        +(n-2)\Big(\frac{\pi^{\frac{n}{2}}}{2\Gamma(\frac{n}{2})}\Big)^{\frac{2}{n-1}}
        -\frac{2\Lambda}{n-1}S^{\frac{2}{n-1}}}\nonumber
\end{eqnarray}
\begin{eqnarray}
        g_{S\Phi}=&&\frac
        {-2(n-1)\Big(\frac{4\Gamma(\frac{n}{2})}{(n-1)\pi^{\frac{n}{2}-1}}\Big)^2
        \Big(\frac{\pi^{\frac{n}{2}}}{2\Gamma(\frac{n}{2})}\Big)^{\frac{2}{n-1}}S}
        {-(n-2)\Big(\frac{4\Gamma(\frac{n}{2})}{(n-1)\pi^{\frac{n}{2}-1}}\Big)^2
        \Big(\frac{\pi^{\frac{n}{2}}}{2\Gamma(\frac{n}{2})}\Big)^{\frac{2}{n-1}}\Phi^2
        +(n-2)\Big(\frac{\pi^{\frac{n}{2}}}{2\Gamma(\frac{n}{2})}\Big)^{\frac{2}{n-1}}
        -\frac{2\Lambda}{n-1}S^{\frac{2}{n-1}}}\nonumber \\
        =&&g_{\Phi S}\nonumber
\end{eqnarray}
\begin{eqnarray}
        g_{\Phi\Phi}=&&\frac
        {-2(n-2)\Big(\frac{4\Gamma(\frac{n}{2})}{(n-1)\pi^{\frac{n}{2}-1}}\Big)^2
        \Big(\frac{\pi^{\frac{n}{2}}}{2\Gamma(\frac{n}{2})}\Big)^{\frac{2}{n-1}}\Phi}
        {-(n-2)\Big(\frac{4\Gamma(\frac{n}{2})}{(n-1)\pi^{\frac{n}{2}-1}}\Big)^2
        \Big(\frac{\pi^{\frac{n}{2}}}{2\Gamma(\frac{n}{2})}\Big)^{\frac{2}{n-1}}\Phi^2
        +(n-2)\Big(\frac{\pi^{\frac{n}{2}}}{2\Gamma(\frac{n}{2})}\Big)^{\frac{2}{n-1}}
        -\frac{2\Lambda}{n-1}S^{\frac{2}{n-1}}}.\nonumber
\end{eqnarray}
Observe that all the metric components have an identical denominator which
appears in the expression of the temperature.

Our concern is the scalar curvature of the Ruppeiner metric, which is
\begin{eqnarray}
        R=\frac{C(S,\Phi)}{A(S,\Phi)B^2(S,\Phi)},\nonumber
\end{eqnarray}
where
\begin{eqnarray}
        A(S,\Phi)&=&-(n-2)\Big(\frac{4\Gamma(\frac{n}{2})}{(n-1)\pi^{\frac{n}{2}-1}}\Big)^2
        \Big(\frac{\pi^{\frac{n}{2}}}{2\Gamma(\frac{n}{2})}\Big)^{\frac{2}{n-1}}\Phi^2\nonumber \\
        &&+(n-2)\Big(\frac{\pi^{\frac{n}{2}}}{2\Gamma(\frac{n}{2})}\Big)^{\frac{2}{n-1}}
        -\frac{2\Lambda}{n-1}S^{\frac{2}{n-1}},
\end{eqnarray}
\begin{eqnarray}
        B(S,\Phi)&=&(n-2)(2n-3)\Big(\frac{4\Gamma(\frac{n}{2})}{(n-1)\pi^{\frac{n}{2}-1}}\Big)^2
        \Big(\frac{\pi^{\frac{n}{2}}}{2\Gamma(\frac{n}{2})}\Big)^{\frac{2}{n-1}}\Phi^2\nonumber \\
        &&-(n-2)\Big(\frac{\pi^{\frac{n}{2}}}{2\Gamma(\frac{n}{2})}\Big)^{\frac{2}{n-1}}
        -\frac{2\Lambda}{n-1}S^{\frac{2}{n-1}},
\end{eqnarray}
and $C(S,\Phi)$ is a regular function whose explicit form is
irrelevant to the singular behavior of $R$. The function $A(S,\Phi)$
is always positive due to the nonextremal condition, as can be seen
from (\ref{tem2}). The function $B(S,\Phi)$ is identical with the
denominator of the heat capacity at constant charge (\ref{hc2}). Hence the
scalar curvature will diverge exactly at those points at which the
heat capacity diverges. It is easy to see that there are two singular points when temperature above $T_c$ and these two points coincide as $T=T_c$, so the thermal geometry method gives the same result which we have found in section III.

%%%%%%%%%%%%%%%%%%%%%%%%%%%%%%%%%%%%%%%%%%%%%%%%%%%%%%%%%%%%%%%%
\section{Discussions}
In the present work, We have obtained different thermodynamic
entities like temperature, potential and heat capacity at constant
charge for a $(n+1)$-dimensional RN-AdS black hole from the first
law of black hole thermodynamics. The heat capacity shows a
divergence at the van der Waals-like critical point. Moreover, we
have investigated the critical behavior of the $(n+1)$-dimensional
RN-AdS black hole at the van der Waals-like critical point. One of
the striking characteristics of the phase transition is the fact
that many measures of a system's behavior near a critical point are
independent of the details of the interactions between the particles
making up the system. The universal features are not only
independent of the numerical details of the interparticle
interactions, but are also independent of the most fundamental
aspects of the structure of the system. The critical exponents of a
$(n + 1)$-dimensional RN-AdS black hole and a van der Waals liquid
gas system are exactly the same. This result is quite interesting
because of the differences in the physical property of the two
systems.

We have also studied the phase transition using the geometrical
perspective of equilibrium thermodynamics. We find that both the
scalar curvature and the heat capacity at constant charge have a
common denominator and hence diverge at identical points. This shows
that a divergence in the scalar curvature corresponds to a
divergence in the heat capacity at constant charge thereby
suggesting the occurrence of a phase transition. Moreover, there is
another factor in the denominator of the scalar curvature which is
the same expression arising in the expression for the temperature
(\ref{tem2}). So we cannot put it equal to zero due to the
nonextremal condition. Therefore we can easily get information about
the occurrence of phase transition from the scalar curvature.

We have found that the van der Waals-like critical behavior
does not present in the planar or hyperbolic RN-AdS black holes. But
we know that there is another type of phase transition associated
with a scalar hair in the planar case \cite{hair}. Based on the
AdS/CFT correspondence, this type of phase transition describes the
superconductivity phase transition in the dual boundary system
(see \cite{review} and references therein), which has been proved a powerful tool to study
superconductivity phenomena. In our case, the boundary system dual
to the spherical RN-AdS black hole should also have critical
behavior dual to the van der Waals-like critical behavior in the
bulk. So, it is quite natural to ask whether this dual boundary
theory also describes some important, realistic phenomena of phase
transition in physics.

We have also observed that RN-dS black holes do not posses the van
der Waals-like phase transition, while the Hawking-Page phase
transition can occur in this kind of background \cite{Carlip}. So,
we see that the asymptotic AdS background is crucial for the van der
Waals-like phase transition. It is interesting to investigate the
underlying mechanism of this phenomenon.

\begin{acknowledgments}
We thank Prof. Y. Ling for helpful discussions. This work is
partly supported by the National Natural Science Foundation of China
(Grant Nos. 10705048, 10731080 and {11075206}) and the President
Fund of GUCAS.
\end{acknowledgments}

\end{document}